\documentclass[journal,onecolumn, 12pt, draftclsnofoot]{IEEEtran}
\usepackage[T1]{fontenc}

\usepackage[font=small]{caption}
\usepackage[labelsep=period]{caption}
\usepackage{float}
\usepackage[latin1]{inputenc}
\usepackage[nospace,noadjust]{cite}
\usepackage{amsfonts}
\usepackage{amsmath}
\usepackage{amsthm}
\usepackage{mathrsfs}
\usepackage{amssymb,mathrsfs}
\usepackage[mathscr]{euscript}
\usepackage{subcaption}
\usepackage{epstopdf} 
\usepackage[dvipsnames]{xcolor}
\usepackage{pgfplots}
\pgfplotsset{compat=newest}
\usetikzlibrary{plotmarks}
\usepackage{amsmath}
\usepackage[normalem]{ulem}
\usepackage{gensymb}

\usepackage[cmintegrals]{newtxmath}

\pgfplotsset{every axis/.append style={
        scaled x ticks = false, 
        x tick label style={/pgf/number format/.cd, fixed, fixed zerofill,
                            int detect, 1000 sep={},precision=3}
    }
}

\DeclareMathOperator*{\argmin}{argmin}

\definecolor{darkpastelgreen}{rgb}{0.01, 0.75, 0.24}

\def\plos{\mathcal{P}_\text{LoS}}

\begin{document}

\title{Aerial Anchors Positioning for Reliable RSS-Based Outdoor Localization in Urban Environments}
\author{\IEEEauthorblockN{
Hazem Sallouha, Mohammad Mahdi Azari, Alessandro Chiumento, Sofie Pollin 
\thanks{Authors are with the Department of Electrical Engineering, KU Leuven, Belgium. Email: h.sallouha@gmail.com
}}}

\maketitle

\begin{abstract}
In this letter, the localization of terrestrial nodes when unmanned aerial vehicles (UAVs) are used as base stations is investigated. Particularly, a novel localization scenario based on received signal strength (RSS) from terrestrial nodes is introduced. In contrast to the existing literature, our analysis includes height-dependent path loss exponent and shadowing  which results in an optimum UAV altitude for minimum localization error. Furthermore, the Cram\'{e}r-Rao lower bound is derived for the estimated distance which emphasizes, analytically, the existence of an optimal UAV altitude. Our simulation results show that the localization error is decreased from over $300\,m$ when using ground-based anchors to $80\,m$ when using UAVs flying at the optimal altitude in an urban scenario.
\end{abstract}
\begin{IEEEkeywords}
Unmanned aerial vehicle (UAV), localization, received signal strength (RSS) , aerial anchors, optimal altitude
\end{IEEEkeywords}

\IEEEpeerreviewmaketitle

\section{Introduction}
Modern connectivity requirements have carried broadband wireless technologies to a new dimension where unmanned aerial vehicles (UAVs) are used as complementary base stations to guarantee connectivity in all accidental circumstances \cite{hourani2,hourani,azari2017ultra,mahdi}. The fast deployment of such connectivity infrastructure has become possible due to recent UAV technology that allows well controlled movement and hovering for small UAVs \cite{fotouhi2017}. Alongside the enormous connectivity potential, location-aware services are also essential, particularly in case of emergencies to locate the user and provide relief services efficiently. 

Localization techniques have been extensively studied in the literature for outdoor scenarios using terrestrial anchors (TAs) \cite{Zanella}. Ranging solutions based on received signal strength (RSS) are particularly attractive because of their intrinsic simplicity and for not requiring extra antennas or time synchronization. A well-known natural representation of the relation between RSS and distance is obtained from the path loss model equation \cite{Zanella}.

UAVs as aerial anchors were presented in \cite{ahmad2,pinotti2016localization,han2016survey}. In \cite{ahmad2} a probabilistic localization algorithm is introduced where a simplified path loss model is used. In \cite{pinotti2016localization} a bound on the position error of terrestrial nodes (TNs) is defined where distances are estimated using round trip time. Moreover, a range-free localization mechanism based on the connectivity information is discussed in \cite{han2016survey}. Nevertheless, all these works ignore the effect of the UAV altitude on both the shadowing effect and the path loss exponent.

To address this issue, we refer to the recent reports that model the channel characteristics for UAV to ground communication links \cite{hourani2, hourani, mahdi,azari2017ultra}. These studies show that the path loss slope and the fluctuation around the average loss due to shadowing (also known as excessive path loss), diminish as the UAV elevation angle and hence altitude increases. The relative variation of these two important factors suggests an interesting trade-off between the lower shadowing effect and smaller path loss exponent at higher altitudes which may finally lead to lower or higher localization error as the UAV altitude varies. To the best of our knowledge, such an important trend in the localization error has not been studied yet and is an open question.

In this paper, we introduce a novel analytical framework for terrestrial nodes (TNs) localization in urban environments using UAVs. In contrast to the previous works, our proposed framework includes height-dependent path loss exponent and shadowing parameters which enables us to explore the optimal positioning of the UAVs for maximum localization accuracy for different types of urban environments. First, by formulating the problem we study the impact of the UAVs altitude and their hight-dependent relative distances for minimum overall localization error. Then, the Cram\'{e}r-Rao lower bound (CRLB) of the estimated distance is derived as a function of both the elevation angle and the UAV-to-TN distance, which affirms analytically the existence of the optimal altitude. Lastly, simulations have been carried out to study the effect of the UAVs altitude, distance between UAVs and the number of UAVs on the localization accuracy. Our results show that the localization error is decreased from over $300\,m$ to $80\,m$ with UAVs at the optimum altitude as compared to that of the TAs in an urban environment. Moreover, the localization error is further decreased with a factor of 2 when optimizing in addition to the altitude, also the distance between the UAVs. Finally, we show that the required number of anchors can be significantly reduced by placing them at altitude.

The rest of the paper is organized as follows. In Section II, we introduce the system model and the path loss model. Then, the localization problem using UAVs is introduced in Section III. In Section IV the simulation results are presented. Finally, the work is concluded in Section V.
\vspace{-1em}
\section{System Model}
\begin{figure}[t]
	\centering	\includegraphics[width=0.55\textwidth,,height=5cm]{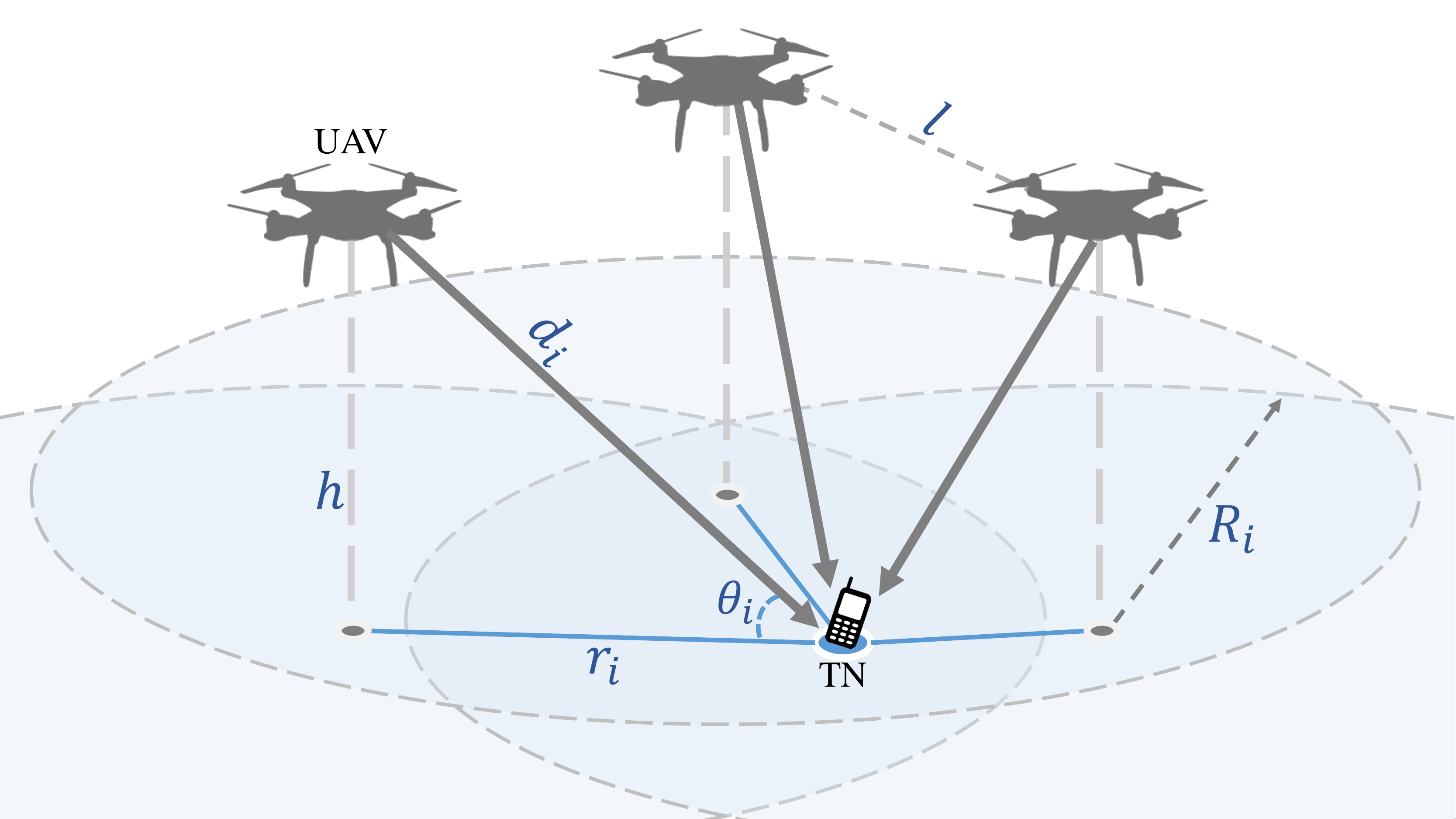} 
	\caption{\small{Localizing TN within the coverage zone of 3 UAVs.}}
	\label{city}
    \vspace{-1.5em}
\end{figure}

Consider a network of UAVs flying within altitudes of $100\,m-3\,km$ and acting as aerial anchors that aim to localize randomly distributed TNs in a given area. The on-board communication technology depends on specific application communication requirements. Appropriate technologies could be as advanced as LTE or WiFi, or as simple as LPWAN \cite{hourani}. To localize TNs, RSS technique is considered, therefore, the conclusion of this paper applies to wide set of technologies \cite{Zanella}. RSS-based ranging techniques require at least three UAVs to localize any TNs. In order to represent the UAVs inter distance with one variable $l$, these UAVs are assumed to fly forming an equilateral triangle with side lengths of $l$ and the same adjustable altitude $h$, as shown in Fig. \ref{city}. As illustrated in this figure, $r_i$ represents the distance between the $i$th UAV's projection and the TN to be localized. The elevation angle of the $i$th UAV with regard to TN is denoted by $\theta_i$.

In order to estimate the distances between UAVs and the TN, a path loss model between the UAV and the TN (in dB) can be used to represent the received power measurements as a function of direct distance $d$ (in meters) and the elevation angle $\theta = \tan^{-1}{(h/r)}$. In this work we consider the path loss model with formulation presented in \cite{goldsmith}. However, we incorporate the dependencies of shadowing and the path loss exponent on the elevation angle as proposed in \cite{hourani2,hourani,mahdi,azari2017ultra}. Subsequently, the path loss model can be written as
\begin{eqnarray}
\text{PL}  = K + 10 \, \alpha(\theta)\log(d/d_o) + \psi(\theta),
\label{ptL}
\end{eqnarray}
where $K$ is a constant dependent on antenna gains and the received power at a reference distance $d_o$ (assumed to be $1\,m$ throughout the paper), $\theta$ is the elevation angle (in radian), and $\alpha(\theta)$ is the path loss exponent. In \eqref{ptL}, moreover, $\psi(\theta)$ represents the shadowing effect which is a log-normal distributed random variable with zero mean and variance $\sigma^{2}(\theta)$ (in dB):
\begin{eqnarray}
\psi(\theta) \sim \mathcal{N}(0,\,\sigma^{2}(\theta)).
\end{eqnarray}
Following \cite{hourani2,hourani,mahdi,azari2017ultra}, one can write
\begin{eqnarray}
\sigma^2(\theta) = \plos^2(\theta)~\sigma^2_{\text{LoS}}(\theta) + [1 - \plos(\theta)]^2 \sigma^2_{\text{NLoS}}(\theta),
\label{sig}
\end{eqnarray}
where $\sigma_{\text{LoS}}(\theta)$ and $\sigma_{\text{NLoS}}(\theta)$ correspond to the shadowing effect of the independently distrusted LoS and non-LoS (NLoS) links respectively, and can be written as
\begin{eqnarray}
\sigma_j(\theta) = a_j \exp{(-b_j\theta)},~~~j \in \{\mathrm{LoS}, \mathrm{NLoS}\}
\label{sigj}
\end{eqnarray}
with $a_j$ and $b_j$ being frequency and environment dependent parameters. Furthermore, $\plos(\theta)$ is the probability of LoS given by
\begin{eqnarray}
\plos(\theta) = \frac{1}{1 + a_o \exp{(-b_o \theta)}},
\label{Prlos}
\end{eqnarray}
where $a_o$ and $b_o$ are environment dependent constants. It is to be noted that $\plos(\theta)$ in (\ref{Prlos}) is an approximation of the $\plos$ given in \cite{hourani} in cases where the height of the TN is disregarded compared to $h$, e.g., for TN's height equal to $1.5\,m$, the minimum altitude, $h^{(min)}$ is $50\,m$. Finally, the path loss exponent is expressed as
\begin{eqnarray}
\alpha(\theta) = a_1 \plos(\theta) + b_1,
\label{pl_expn}
\end{eqnarray}
where $a_1 = \alpha(\frac{\pi}{2}) - \alpha(0)$ and $b_1 = \alpha(0)$. Accordingly, $\alpha(\theta)$ decreases as $\theta$ increases within the bounds $\alpha(\frac{\pi}{2}) \leq \alpha(\theta) \leq \alpha(0)$. In particular, $\alpha(\frac{\pi}{2})$ and $\alpha(0)$ are the path loss exponents for free space and the adopted environment respectively. For instance, $\alpha(\frac{\pi}{2}) = 2$ and $\alpha(0) \in [2.7,3.5]$ in urban areas \cite{goldsmith}. Consequently, for urban areas, $2 \leq \alpha(\theta) \leq 3.5$.
\vspace{-1em}
\section{Localization Using Aerial Anchors}
The position of a TN in 2D coordinates is defined as $(x, y)$. Given the fixed projection $(x_i, y_i)$ of the $i$th UAV on the ground, we first estimate the distances $r_i = {\sqrt{(x - x_i)^2+(y-y_i)^2}}$. Subsequently, the multilateration (trilateration in case of three anchors) method can be used to estimate the node's position. In multilateration, least squares are used to estimate the location of the node ($\hat{x}, \hat{y}$) according to the estimated distances.
\vspace{-1em}
\subsection{Problem Formulation}
Assuming that $R_i$ is the coverage radius of the $i$th UAV as illustrated in Fig. \ref{city}, in order to localize a TN it has to be within the coverage zone of at least three UAVs (i.e., $r_i$ $\leq$ $R_i$ where $i = 1, 2, ..., N$ and $N$ $\geq$ 3 is the number of available UAVs). Consequently, the first constraint to localize the node is expressed as $r_i$ $\leq$ $R_i$. Now given the estimated distance $r_i$ and known projection ($x_i$, $y_i$) of the $i$th UAV, the position ($x$, $y$) of the TN can be estimated by finding the ($\hat{x}, \hat{y}$) that satisfies
\begin{eqnarray}
(\hat{x}, \hat{y}) = \argmin_{x,y} \bigg\{\sum_{i=1}^{N} \Big( \sqrt{(x - x_i)^2+(y-y_i)^2} - {\hat{r}}_i \Big)^2 \bigg\},
\label{opt1}
\end{eqnarray}
where ${\hat{r}}_i = \sqrt{{\hat{d}}_i^2 - h^2}$. This formulation provides the nodes' positions for a given $h$. For an estimated location ($\hat{x}, \hat{y}$) of a node, the localization error is given by
\begin{eqnarray}
\xi = \hspace{0.1cm} \parallel {\hat{\boldsymbol{r}}} - \boldsymbol{r} \parallel \hspace{0.1cm} =  \sqrt{\sum_{i=1}^{N} \mid\hat{r}_i - r_i \mid^2},
\label{locoErr}
\end{eqnarray}
where $\boldsymbol{r} = [r_1, r_2, ..., r_N]$, ${\hat{\boldsymbol{r}}} = [\hat{r}_1, \hat{r}_2, ..., \hat{r}_N]$ and $\lVert.\lVert$ represents the euclidean distance. For simplicity one can assume that the TN is in the range of the $i$th UAV for all examined values of $h$. Now, in order to find the optimal altitude $h^{(opt)}$ that minimizes the localization error, the optimization problem can be described as
\begin{equation}
\begin{aligned}
& \underset{h \in [h^{(min)}, \infty)}{\text{minimize}}
& & {\{\xi\}} \\
& \text{subject to}
& & {\hat{r}}_i = \sqrt{{\hat{d}}_i^2 - h^2}, \\
&&& r_i \leq R_i.
\label{optProb}
\end{aligned}
\end{equation}
In (\ref{optProb}) we minimize the error for $h$. The optimization problem depends explicitly on $N$ and implicitly on $l$, as changing $l$ affects both the horizontal distance $r_i$ and the direct distance $d_i$. Intuitively, the estimated distance ${\hat{d}}_i^2$ between a TN and the $i$th UAV depends on the path loss model, and hence on the height of the UAV. Referring to (\ref{sig}), for a given TN the shadowing effect at low values of $h$ will be relatively high causing large localization error. As $h$ increases, the shadowing effect decreases: based on (\ref{pl_expn}) concurrently, the resolution will also decrease. In the case of low resolution namely, low slope, small shadowing will produce a large localization error. Therefore, we assert that for a group of TNs, an optimum altitude $h$ that minimizes the localization error is likely to be present. The existence of such altitude in which the effect of lower resolution and shadowing is balanced, is numerically studied in Section IV.
\vspace{-1em}
\subsection{Cram\'{e}r-Rao lower bound}
A crucial performance benchmark for ranging-based localization techniques is the CRLB of the estimated distance as it well defines a lower limit for the variance of any (asymptotically) unbiased estimator \cite{van2004detection}. To obtain the CRLB for the estimated distance, recall the path loss model given in (\ref{ptL}). This model describes the behavior of the received power at different distances and altitudes. Additionally, it considers the randomness due to the shadowing. Consequently, the time-averaged received power (in dBm) can be expressed as
\begin{eqnarray}
\mathcal{P}_{r} = - 10 \alpha(\theta)\log(d) - K - \psi(\theta) + C,
\label{prx}
\end{eqnarray}
where C is a constant (in dBm) which depends on the transmit
power and received power to RSS transduction. Note that for a given distance $d$, the time-averaged received power given in (\ref{prx}) is a stochastic variable following a shifted version of the probability density function (PDF) of $\psi(\theta)$. Accordingly, the PDF of $\mathcal{P}_{r}$ conditioned on $d$ and $\theta$ is given by
\begin{eqnarray}
f_{\mathcal{P}_{r}|d,\theta}(w) = f_{\psi | d,\theta}( -w - 10 \alpha(\theta)\log(d) - K + C),
\label{Pr-pdf}
\end{eqnarray}
where $w$ ia an auxiliary variable. The CRLB of the estimated distance denoted as $\hat{d}$ is then expressed as \cite{van2004detection}
\begin{eqnarray}
\sigma_{\text{CRLB}}^2 \geq \frac{1}{E_{\mathcal{P}_{r}|d,\theta}\bigg\{ \bigg[ \frac{\partial}{\partial d} \ln f_{\mathcal{P}_{r}|d,\theta}(w) \bigg]^2 \bigg\}},
\label{cr}
\end{eqnarray}
where $E_{\mathcal{P}_{r}|d,\theta}\{.\}$ is the expectation conditioned on $d$ and $\theta$. Using the PDF given in (\ref{Pr-pdf}), it is straightforward to show that
\begin{eqnarray}
\frac{\partial}{\partial d} \ln f_{\mathcal{P}_{r}|d,\theta}(w) &=& [-w - 10~ \alpha(\theta)\log(d) - K + C] \nonumber\\
&\times& \frac{10~\alpha(\theta)}{d~\ln(10)~\sigma^2(\theta)}
\label{crlbDiff}
\end{eqnarray}
We now proceed to write a closed-form equation by substituting (\ref{sig}) and (\ref{pl_expn}) in (\ref{crlbDiff}) and subsequently (\ref{crlbDiff}) in (\ref{cr}). Then, after simplifications, one obtains
\begin{eqnarray}
{\sigma}_{\text{CRLB}} &\geq&  \frac{d \, \ln (10)}{10} \,\, \label{crlb-f} \\ &\times&\frac{\sqrt{\plos^2(\theta)~\sigma^2_{\text{LoS}}(\theta) + [1 - \plos(\theta)]^2 \sigma^2_{\text{NLoS}}}}{  a_1 \plos(\theta) + b_1 }. \nonumber
\label{crlFinal}
\end{eqnarray}
It is worth noting that, ${\sigma}_{\text{CRLB}}$ given in (\ref{crlb-f}) presents a performance measure for individual range estimators, i.e., UAVs. Evidently, from (\ref{crlb-f}), the accuracy of the estimated distance depends on both $d$ and $\theta$. Therefore, one is able to control the system performance by adjusting the elevation angle which, for a given $r$, implicitly adjusts the UAV's altitude $h$. In general, the existence of an estimator that achieves the CRLB confirms the tightness and hence the benefit of the CRLB \cite{van2004detection}. In the next section we investigate the maximum likelihood estimator (MLE) and show that it resembles the CRLB.
\section{Simulation Results and Discussion}
We assume 1000 TNs uniformly distributed in a circular area with a radius of $1000\,m$ and centered at ($x$, $y$)  = (0, 0). Moreover, a system frequency of 2GHz is considered. The path loss model considered in this work has some constants that control the shadowing ($a_j, b_j$) with j $\in$ $\{\mathrm{LoS},\mathrm{NLoS}\}$, the probability of LoS ($a_o, b_o$) and the path loss exponent ($a_1,b_1$). The values used for these constants are summarized in Table \ref{const}. These values were chosen as recommended in \cite{hourani2, hourani,azari2017ultra, goldsmith} for urban environments. In the following, we assumed three UAVs placed as vertices of an equilateral triangle with sides length of $l$ = $500\,m$ are serving the TNs, unless mentioned otherwise. Each UAV is assumed to collect five RSS samples at the same position before estimating its distance to the TN.
\begin{table}[t]
	\caption{\small{The constants used in the simulation results.}}
    \vspace{-1.5em}
	\begin{center}
		\begin{tabular}{ | l || c | c |}
			\hline
			Parameter & Suburban & Urban \\
			\hline
			\hline
			$a_{\text{LoS}}$ & 5 & 10\\
			\hline
			$b_{\text{LoS}}$ &  3.5 & 2.5 \\
			\hline
			$a_{\text{NLoS}}$ & 10 & 30\\
			\hline
			$b_{\text{NLoS}}$ &  2.5 & 1.7 \\
			\hline
			$a_o$ &  47 & 45 \\
			\hline
			$b_o$ &  20 & 10 \\
			\hline
			$a_1$ & -1 & -1.5 \\
			\hline
			$b_1$ &  3 & 3.5 \\
			\hline
		\end{tabular}
	\end{center}
	\label{const}
    \vspace{-1em}
\end{table}

\begin{figure}[t]
	\centering
	\input{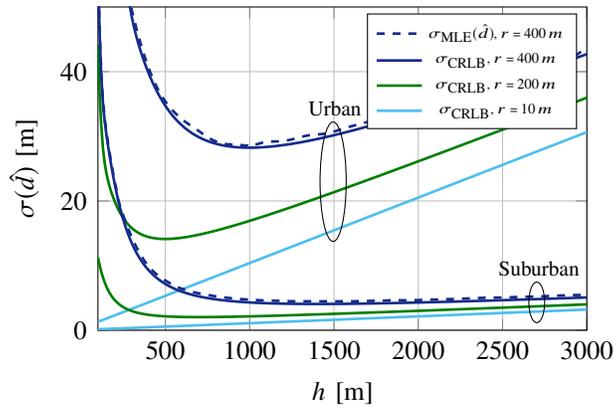}
    \vspace{-0.8em}
	\caption{\small{$\sigma_{\text{MLE}}(\hat{d})$ and $\sigma_{\text{CRLB}}$ of the estimated distance versus $h$.}}
	\label{varr}
    \vspace{-1.5em}
\end{figure}

\textbf{CRLB trend and its tightness}: In Fig. \ref{varr} the CRLB of the estimated distance between one UAV and one TN, as given in (\ref{crlb-f}), is presented as a function of $h$. Firstly, the standard deviation of MLE, $\sigma_{\text{MLE}}(\hat{d})$ as compared to $\sigma_{\text{CRLB}}$ is illustrated. As shown in the figure, $\sigma_{\text{MLE}}(\hat{d})$ resembles $\sigma_{\text{CRLB}}$ which confirms the tightness of $\sigma_{\text{CRLB}}$. Secondly, the figure shows that for large values of $r$, $\sigma_{\text{CRLB}}$ first decreases and then increases: 1) $\sigma_{\text{CRLB}}$ decreases with $h$: although $d$ and $\theta$ increase with $h$, a slight increase of $\theta$ will exponentially increase $\plos$. 2) the CRLB increases with $h$, as $h \gg r$, a small increase of $h$ will considerably increase $d$ making it the dominant part. Finally, for small values of $r$, e.g., $r$ = $10\,m$, $\sigma_{\text{CRLB}}$ represents a linear function of $h$. When $r$ = $10\,m$, increasing $h$ makes $\plos$ converges to one rapidly. Consequently, increasing $h$ will equivalently increase $d$, producing a similar behavior as in TAs' case.

\begin{figure}
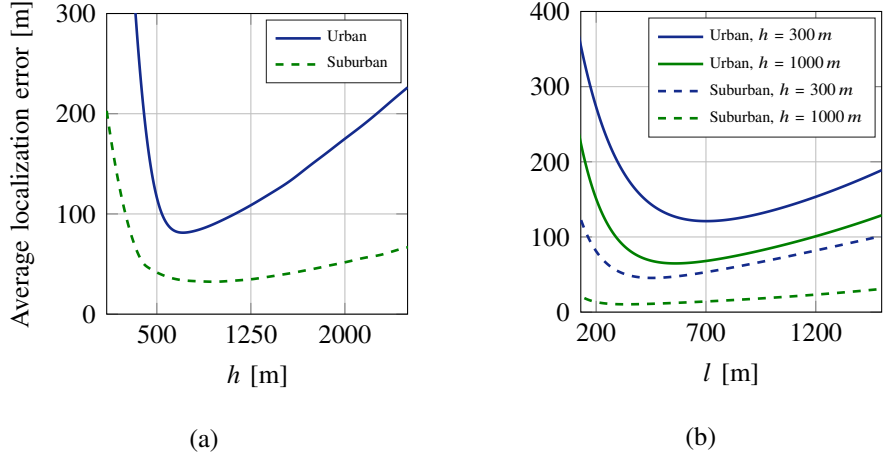

	\centering
	\begin{subfigure}{.4\textwidth}
		\centering
		\input{LocErr.tex}
        \caption{}
		\label{LocErr}
	\end{subfigure}%
	\begin{subfigure}{.4\textwidth}
		\centering
		\input{d-UAV.tex}
		\caption{}
		\label{duav}
	\end{subfigure}
    \vspace{-0.5em}
	\caption{\small{(a) Average localization error over uniformly distributed TNs versus $h$. (b) Localization error versus inter distance $l$ for a TN $650\,m$ away from the center of the network.}}
	\label{fig:test}
	\vspace{-1.5em}
\end{figure}

\textbf{The impact of UAVs altitude}: After investigating the individual distance estimator in Fig. \ref{varr}, now we proceed to check the average localization error when three UAVs are used. The average localization error for all TNs in the considered network as a function of $h$ is shown in Fig. \ref{LocErr}. It can be concluded, from the figure, that in urban areas there exist a global minimum at $h$ $\approx$ $750\,m$ for the average error over $\boldsymbol{r}$. For a deeper look, consider the altitudes less than $750\,m$, here the localization error decreases as $h$ increases, since the shadowing effect will be decreasing with $h$. Moreover, $\plos$ is also increasing with $h$. On the other hand, when $h$ exceeds $750\,m$, the low resolution curve will be more sensitive to shadowing effects (i.e., small variations in the path loss model curve will lead to a large estimation error), this leads to an inverse relation between $h$ and the localization accuracy. This interesting trade off between the curve resolution, represented by the path loss exponent and the shadowing, produces $h^{(opt)}$ at which the two factors are balanced. Same trade-off also exists in the suburban environments as shown in Fig. \ref{LocErr}. 
In suburban areas an improvement over $150\,m$ on localization accuracy is obtained, whereas in case of urban environments we can go from over $300\,m$ localization error using TAs to $80\,m$ using UAVs flying at $h^{(opt)}$. It is worth remarking that, the optimum altitude $h^{(opt)}$ can be mapped to an optimal evaluation angle as for an average distance $\overline{r}$, $\theta^{(opt)} = \tan^{-1}(h^{(opt)}/\overline{r})$. For instance, this gives us $\theta^{(opt)} \approx 50^{\circ}$ in urban areas. 

\textbf{The impact of UAVs inter distances}: Fig. \ref{duav} presents the localization error for a TN $650\,m$ (the average distance from the center) away from the center of the network as a function of the distance between UAVs for urban and suburban environments. As shown in the figure, for urban areas, at an initial UAV distance of $l = 100\,m$ the localization error reaches $200\,m$ for $h$ = $1000\,m$. This is due to the fact that, in trilateration method, at small $l$ any low estimation error in the distance will lead to a large error in the estimated location. Moreover, increasing $l$ improves the localization accuracy where an accuracy of $60\,m$ is achieved at $l$ = $600\,m$ for the same $h$. Eventually, as $l$ keeps increasing, the distance will become too large to give an acceptable distance estimate, making a low localization error impossible.

\begin{figure}[t]
	\centering
	\input{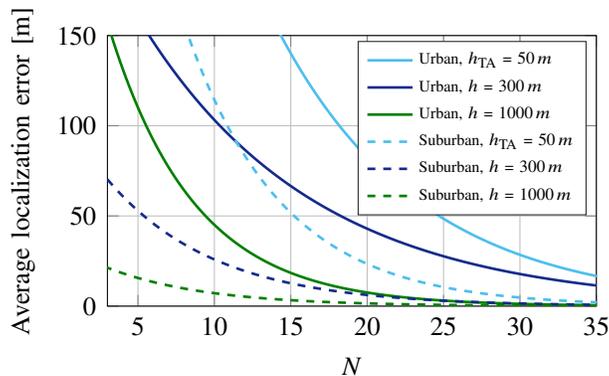}
    \vspace{-0.8em}
	\caption{\small{Localization error versus number of UAVs for a TN $650\,m$ away from the center of the network.}}
	\label{Nuav}
    \vspace{-1.5em}
\end{figure}

\textbf{The impact of the number of UAVs}: The localization accuracy with different number of UAVs with increment of 3 (i.e., $N$ = 3, 6, 9, etc) is presented in Fig. \ref{Nuav}. We consider adding UAVs in equilateral triangles with increasing edge $l$ of $20\,m$ steps, i.e., if $l$ = $100\,m$ for $N$ = 3, then the next triangle has $l$ = $120\,m$. As shown in the figure, the impact of adding more UAVs as aerial anchors can significantly increase the localization accuracy, even without optimizing $h$ or $l$. As one can see, the $150\,m$ accuracy that can be achieved with three UAVs at $h$ = $1000\,m$, requires around 15 TAs with a typical height $h_{\text{TA}}$ = $50\,m$. Accordingly, to achieve a predefined localization error, the number of required aerial anchors is much lower than that of TAs which highlights the effectiveness of UAVs deployment for localization. 
\vspace{-0.5em}
\section{Conclusion}
\vspace{-0.2em}
The improvement in localization accuracy of TNs when using UAV anchors and relying on RSS techniques has been investigated. Particularly, we defined an optimal altitude at which the localization error of TNs is minimized. Furthermore, CRLB of the estimated distance has been derived confirming the impact of the altitude on the distance estimation error. Our study confirmed the effectiveness of placing anchors at altitude in terms of both localization error and the required number of anchors for a target accuracy. Finally, the reported results motivate investigating other localization techniques, such as angle-of-arrival, and potentially integrate them with the proposed RSS-based framework for further improvements.

\vspace{-1em}
\section*{Acknowledgment}
The authors would like to thank Prof. Hugo Van Hamme for the discussions about maximum likelihood estimator. This work was partially funded by the KUL OT Capacities.

\vspace{-0.5em}
\bibliographystyle{ieeetr} 
\bibliography{bib}

\end{document}